\documentclass[12pt,prb]{revtex4}

\usepackage{amsmath}

\begin{document}

\newcommand{\vv}{\mathbf{v}}
\newcommand{\uv}{\mathbf{u}}
\newcommand{\nv}{\mathbf{n}}
\newcommand{\nhat}{\hat{\nv}}
\newcommand{\uhat}{\hat{\uv}}

\title{Direct simulation for a homogeneous gas}

\author{Hasan Karabulut}

\affiliation{Rize University,
Faculty of Arts and Sciences, Department of Physics,
53100 Rize Turkey}

\begin{abstract}
A probabilistic analysis of the direct simulation of a homogeneous gas
is given. A hierarchy of equations similar to the BBGKY hierarchy
for the reduced probability densities is derived. By invoking the molecular
chaos assumption, an equation similar to the Boltzmann equation for
the single particle probability density and the corresponding
H-theorem is derived.
\end{abstract}

\maketitle 

\section{Introduction}

Direct simulation Monte Carlo method (DSMC) is a standard method for solving
the Boltzmann equation numerically. In this method space is divided into
cells of volume $\Delta V$ and a large number of ``particles'' ($N=10^{3}$--$%
10^{6}$) represent the real gas molecules. The evolution of the gas for a
short time $\Delta t$ is calculated in two steps. In the first step all
particles are propagated for a time $\Delta t$ without collisions. In the
second step some randomly chosen pairs of particles in the same cell are
allowed to collide and change their velocities without changing their
positions. Number of pairs $(n)$ chosen to make collision attempts is given
by the formula $n=RN^{2}\Delta t/2V$ where $R$ is a parameter we choose, $N$
is the number of particles in the cell and $V$ is the volume of the cell. We
call $n\,$number of collision attempts because not every chosen pair makes a
collision. A pair is allowed to make a collision with a probability $u\sigma
_{T}/R$, where $u$ is their relative velocity and $\sigma _{T}$ is the total
cross section. The results are not sensitive to value of the parameter $R$
as long as it is big enough such that very few pairs violate the condition $%
u\sigma /R\leq 1$ since average number of successful collision attempts
\[
n\frac{\left\langle u\sigma _{T}\right\rangle }{R}=\frac{RN^{2}\Delta t}{2V}%
\frac{\left\langle u\sigma _{T}\right\rangle }{R}=\frac{N^{2}\left\langle
u\sigma _{T}\right\rangle }{2V}\Delta t,
\]
is independent of $R$. Here $\left\langle u\sigma _{T}\right\rangle $ is the
average of $u\sigma _{T}$ over all possible pairs. Although taking a very
big $R$ is acceptable theoretically, for practical reasons $R$ should not
chosen be too big either.

The original method is due mainly to G. A. Bird. A seminal paper\cite{Bird70}
of Bird gave some heuristic arguments to justify its use. There are many
good references on the subject. Ref. \onlinecite{Garcia} has a good tutorial
on the subject and Ref. \onlinecite{Bird94} is a monograph on the subject by
Bird himself which is a complete reference for the developments up to its
publication year 1994. Also books on rarified gas dynamics devote many
chapters to the subject and Ref. \onlinecite{Carlo} and Ref. %
\onlinecite{Shen} are useful references in this category.

A variant of the method was derived by Nanbu\cite{Nanbu80} starting from the
Boltzmann equation. To represent the evolution of the real gas such methods
should converge to the true solution of the Boltzmann equation in the limit $%
N\rightarrow \infty $, $\Delta V\rightarrow 0$, and $\Delta t\rightarrow 0$.
Convergence proofs were given by Babovsky\cite{Babovsky1} and Babovsky and
Illner\cite{Babovsky2} for Nanbu's method and by Wagner\cite{Wagner92} for
Bird's method.

For the evolution of the velocity distribution of a spatially uniform gas
there is no need to divide physical space into cells and we can just work in
velocity space. Although Bird recommended\cite{Bird94} dividing real space
into cells for studying the spatially homogeneous gas, we will show that
this division is unnecessary. If we consider velocity space only and collide
random pairs, we should obtain the evolution of the velocity distribution.
The purpose of this paper is to study this stochastic process.

These efforts to solve the Boltzmann equation using stochastic methods were
driven by scientific applications and there was no motivation to use them as
a pedagogical tool. It is surprising that similar stochastic algorithms for
the homogeneous gas were conceived by people interested in using them as a
pedagogical tool to demonstrate the evolution of a gas to the
Maxwell-Boltzmann distribution. The earliest of such articles of which the
author is aware is that of Novak-Bortz\cite{Novak70} who studied the
evolution of a gas of two-dimensional disks. Their algorithm is based on
taking random pairs and colliding them with a probability proportional to $%
u\sigma$. Eger and Kress\cite{Eger82} modified this algorithm and Bonomo and
Riggi\cite{Bonomo84} applied the modification to hard disks. There are also
other papers\cite{Sauer81,Berger88} that do not use DSMC type stochastic
processes to demonstrate the Maxwell-Boltzmann distribution. Although the
DSMC method was well known, these papers do not reference papers on DSMC.
Apparently the idea of stochastic methods for the evolution of a homogeneous
gas was conceived for pedagogical applications independently.

As mentioned, the DSMC algorithm can be used to demonstrate the approach of
a velocity distribution to the Maxwell-Boltzmann distribution. The algorithm
also gives an estimate of how many collisions is required to reach
equilibrium and how various parameters affect the evolution of the system.

Although direct simulation is intuitively appealing, it is not clear that
direct simulation algorithms represent the evolution of a real gas. The
convergence proofs we have cited are formal and difficult to read. In this
paper we prove that in the direct simulation appropriately normalized single
particle probability distribution satisfies the Boltzmann equation for a
homogeneous gas. The proof is relatively easy and intuitively appealing and
also its language is familiar to the physicist from the well known BBGKY
hierarchy.\cite{Huang}

In Sec.~II we consider the stochastic algorithm for a homogeneous gas. We
derive a hierarchy of equations for the probability distribution of
particles similar to the BBGKY hierarchy.\cite{Huang} We use the molecular
chaos assumption to derive an equation similar to the Boltzmann equation for
the single particle probability distribution $f(\mathbf{v})$. We derive an
H-theorem for $f(\mathbf{v})$ and prove convergence to equilibrium. We also
show how the equation for $f(\mathbf{v})$ reduces to the Boltzmann equation
for a particular choice of collision probabilities and derive Bird's ``time
counter'' and ``no time counter'' methods.

\section{Analysis of the direct simulation algorithm for a homogeneous gas}

Consider a homogeneous gas of $N\gg 1$ molecules without internal degrees of
freedom. We randomly select pairs of molecules to collide. All possible
pairs have an equal probability of $2/(N-1)N$ to be selected. Suppose the
velocities of the pair are $\mathbf{v}_{A}$ and $\mathbf{v}_{B}$. The
conditional probability that after the collision they have the velocities $%
\mathbf{v}_{C}$ and $\mathbf{v}_{D}$ in the intervals $d^{3}\mathbf{v}_{C}$
and $d^{3}\mathbf{v}_{D}$ is $T(\mathbf{v}_{A},\mathbf{v}_{B};\mathbf{v}_{C},%
\mathbf{v}_{D})\,d^{3}\mathbf{v}_{C}\,d^{3}\mathbf{v}_{C}$. (From now on we
will denote $d^{3}\mathbf{v}$ as $d\mathbf{v}$ for simplicity.) We also
assume the symmetries
\begin{subequations}
\label{eq1}
\begin{align}
T(\vv_{A},\vv_{B};\vv_{C},\vv_{D})
&= T(\vv_{C},\vv_{D};\vv_{A},\vv_{B}) \label{a10} \\
T(\vv_{A},\vv_{B};\vv_{C},\vv_{D})
&= T(\vv_{B},\vv_{A};\vv_{D},\vv_{C}). \label{a20}
\end{align}
\end{subequations}
The total probability is unity and therefore
\begin{equation}
\int \!T(\mathbf{v}_{A},\mathbf{v}_{B};\mathbf{v}_{C},\mathbf{v}_{D})\,d%
\mathbf{v}_{C}\,d\mathbf{v}_{D}=\!\int \!T(\mathbf{v}_{A},\mathbf{v}_{B};%
\mathbf{v}_{C},\mathbf{v}_{D})\,d\mathbf{v}_{A}\,d\mathbf{v}_{B}=1.
\label{a40}
\end{equation}
Every selected pair makes a collision, although as we will show, by defining
$T(\mathbf{v}_{A},\mathbf{v}_{B};\mathbf{v}_{C},\mathbf{v}_{D})$ some of the
collisions do not change velocities. After each collision new velocities of
the molecules are replaced by the old ones, and we select a new pair for the
next collision. Of course there is the possibility of choosing the same pair
with a very small probability. If that happens we let them collide again. We
don't keep record of pairs that have made collisions already.

We define $f^{(N)}(\mathbf{v}_{1},\mathbf{v}_{2},\ldots ,\mathbf{v}_{N})$ as
the probability density for the molecules. Because the molecules are
indistinguishable, we require that the $f^{(N)}$ be totally symmetric:
\begin{equation}
f^{(N)}(\mathbf{v}_{1},\ldots ,\mathbf{v}_{i},\ldots ,\mathbf{v}_{j},\ldots ,%
\mathbf{v}_{N})=f^{(N)}(\mathbf{v}_{1},\ldots ,\mathbf{v}_{j},\ldots ,%
\mathbf{v}_{i},\ldots ,\mathbf{v}_{N}).  \label{a50}
\end{equation}
We also define the reduced probability densities
\begin{equation}
f^{(M)}(\mathbf{v}_{1},\mathbf{v}_{2},\ldots ,\mathbf{v}_{M})=\!\int
\!f^{(N)}(\mathbf{v}_{1},\mathbf{v}_{2},\ldots ,\mathbf{v}_{N})\,d\mathbf{v}%
_{M+1}\,d\mathbf{v}_{M+2}\ldots d\mathbf{v}_{N}.  \label{a60}
\end{equation}
Because we will be dealing with pairs of particles, it is useful to define
\begin{equation}
f_{i,j}^{(M)}(\mathbf{v}_{A},\mathbf{v}_{B})=f^{(M)}(\mathbf{v}_{1},\ldots ,%
\mathbf{v}_{i}=\mathbf{v}_{A},\ldots ,\mathbf{v}_{j}=\mathbf{v}_{B},\ldots ,%
\mathbf{v}_{M}).  \label{a70}
\end{equation}
That is, the velocities of the $i,j$ pair are replaced by $\mathbf{v}_{A},%
\mathbf{v}_{B}$ in the $f^{(M)}(\mathbf{v}_{1},\mathbf{v}_{2},\ldots ,%
\mathbf{v}_{M})$ where $i,j\leq M$. We will also use the notation $f^{(M)}(%
\mathbf{v};n)$ for $f^{(M)}(\mathbf{v}_{1},\mathbf{v}_{2},\ldots ,\mathbf{v}%
_{M})$ after the $n^{th}$ collision.

The function $f^{(N)}(\mathbf{v};n)$ satisfies the equation
\begin{equation}
f^{(N)}(\mathbf{v};n+1)=\frac{1}{N(N-1)}\sum_{i=1}^{N}\sum_{j\neq
i}^{N}\!\int\! f_{i,j}^{(N)}(\mathbf{v}_{A},\mathbf{v}_{B};n)T(\mathbf{v}%
_{A},\mathbf{v}_{B};\mathbf{v}_{i},\mathbf{v}_{j}) \,d\mathbf{v}_{A}\,d%
\mathbf{v}_{B}.  \label{a80}
\end{equation}
The meaning of Eq.~\eqref{a80} is clear. If $i,j$ is the last pair of
molecules that has collided, then the probability of having $\mathbf{v}_{i},%
\mathbf{v}_{j}$ pairs after the collision is the probability of having
initial velocities $\mathbf{v}_{A},\mathbf{v}_{B}$ (represented by $%
f_{i,j}^{(N)}(\mathbf{v}_{A}, \mathbf{v}_{B})$) multiplied by the
probability of ending with $\mathbf{v}_{i},\mathbf{v}_{j}$ (represented by $%
T(\mathbf{v}_{A},\mathbf{v}_{B}; \mathbf{v}_{i},\mathbf{v}_{j})$). The sum
over $i,j$ and the factor $1/N(N-1)$ in Eq.~\eqref{a80} represents the fact
that all pairs are possible with probability $1/N(N-1)$.

If we integrate Eq.~(\ref{a80}) over $\mathbf{v}_{M+1},\mathbf{v}%
_{M+2},\ldots, \mathbf{v}_{N}$, we obtain
\begin{eqnarray}
f^{(M)}(\mathbf{v};n+1) &=&\frac{(N-M)(N-M-1)}{N(N-1)}f^{(M)}(\mathbf{v};n)
\nonumber \\
&&{}+\frac{2(N-M)}{N(N-1)}\sum_{i=1}^{M}\!\int\! f_{i,M+1}^{(M+1)}(\mathbf{v}%
_{A}, \mathbf{v}_{B};n)T(\mathbf{v}_{A},\mathbf{v}_{B};\mathbf{v}_{i},%
\mathbf{v}_{M+1})\,d\mathbf{v}_{A}d\mathbf{v}_{B}\,d\mathbf{v}_{M+1}
\nonumber \\
&&{}+\frac{M(M-1)}{N(N-1)}\sum_{i=1}^{M}\sum_{j\neq i}^{M}\!\int\!
f_{i,j}^{(M)}(\mathbf{v}_{A},\mathbf{v}_{B};n) T(\mathbf{v}_{A},\mathbf{v}%
_{B};\mathbf{v}_{i},\mathbf{v}_{j})\,d\mathbf{v}_{A}\,d\mathbf{v}_{B}.
\label{a90}
\end{eqnarray}
The $f^{(M)}(\mathbf{v};n+1)$ depends on $f^{(M+1)}(\mathbf{v};n)$; Eq.~%
\eqref{a90} represents a hierarchy of equations similar to the BBGKY
hierarchy.\cite{Huang}

The first equation in the hierarchy is
\begin{eqnarray}
f^{(1)}(\mathbf{v};n+1) &=&(1-2/N)f^{(1)}(\mathbf{v};n)  \nonumber \\
&&{}+\frac{2}{N}\!\int \!f^{(2)}(\mathbf{v}_{A},\mathbf{v}_{B};n)T(\mathbf{v}%
_{A},\mathbf{v}_{B};\mathbf{v}_{C},\mathbf{v})d\mathbf{v}_{A}\,d\mathbf{v}%
_{B}\,d\mathbf{v}_{C}.  \label{a100}
\end{eqnarray}
If we make the assumption of molecular chaos
\begin{equation}
f^{(2)}(\mathbf{v}_{A},\mathbf{v}_{B};n)=f^{(1)}(\mathbf{v}_{A};n)f^{(1)}(%
\mathbf{v}_{B};n),  \label{a110}
\end{equation}
we obtain a nonlinear equation for $f^{(1)}(\mathbf{v};n)$ similar to the
Boltzmann equation. For large $N$ this approximation is almost exact as
shown by the following argument. The velocities $\mathbf{v}_{1},\mathbf{v}%
_{2}$ can be correlated only if particles one and two have collided with
each recently. But this probability is of order $1/N$, which implies that
for large $N$, the velocity distributions of any two particles are
uncorrelated. Present personal computers can handle $N=10^{5}$--$10^{6}$ so
the assumption is almost exact.

The key assumption in the argument for the validity of Eq.~\eqref{a110} is
``recently.'' Two particles might be correlated for a short time, but after
they have made a few collisions with other particles the correlations are
expected to disappear.

Another simplification occurs for large $N$. The factor of $2/N$ in Eq.~(\ref
{a100}) is small and thus we can consider $\tau =2n/N$ to be a continuous
parameter which we call the collision time. Then $\Delta \tau =2/N$ and $%
\big[ f^{(1)}(\mathbf{v};n+1)-f^{(1)}(\mathbf{v};n)\big]/\Delta \tau$ can be
written as $\partial f^{(1)}(\mathbf{v};\tau)/\partial \tau$ and Eq.~(\ref
{a100}) becomes
\begin{equation}
\frac{\partial f^{(1)}(\mathbf{v};\tau)}{\partial \tau}=-f^{(1)}(\mathbf{v}%
;\tau)+\!\int\! f^{(1)}(\mathbf{v}_{A};\tau)f^{(1)}(\mathbf{v}_{B};\tau)T(%
\mathbf{v}_{A}, \mathbf{v}_{B};\mathbf{v}_{C},\mathbf{v})\,d\mathbf{v}%
_{A}\,d \mathbf{v}_{B}\,d\mathbf{v}_{C}.  \label{a120}
\end{equation}
From now on we will suppress the superscript $(1)$ and the collision time $%
\tau$ in $f^{(1)}(\mathbf{v};\tau)$. Equation~\eqref{a120} can be expressed
as
\begin{equation}
\frac{\partial f(\mathbf{v})}{\partial \tau}=-f(\mathbf{v}) + \!\int\! f(%
\mathbf{v}_{A})f(\mathbf{v}_{B})T(\mathbf{v}_{A},\mathbf{v}_{B};\mathbf{v}%
_{C}, \mathbf{v})\,d\mathbf{v}_{A}\,d\mathbf{v}_{B}\,d\mathbf{v}_{C}.
\label{a130}
\end{equation}

\subsection{The H-theorem and approach to equilibrium}

By using the relation
\begin{equation}
f(\mathbf{v})=\!\int\! f(\mathbf{v})f(\mathbf{v}_{C})T(\mathbf{v}_{A},%
\mathbf{v}_{B};\mathbf{v}_{C},\mathbf{v})\,d\mathbf{v}_{A}\,d\mathbf{v}%
_{B}d\,\mathbf{v}_{C},  \label{b10}
\end{equation}
which follows from Eq.~(\ref{a40}) and the normalization of $f(\mathbf{v}%
_{C})$, we can write Eq.~(\ref{a130}) as
\begin{equation}
\frac{\partial f(\mathbf{v})}{\partial \tau} =\!\int\! \big[ f(\mathbf{v}%
_{A})f(\mathbf{v}_{B})-f(\mathbf{v})f(\mathbf{v}_{C})\big] T(\mathbf{v}_{A},
\mathbf{v}_{B};\mathbf{v}_{C},\mathbf{v})\,d\mathbf{v}_{A}\,d\mathbf{v}%
_{B}\,d \mathbf{v}_{C}.  \label{b20}
\end{equation}
This form is similar to the Boltzmann equation.

We can derive an H-theorem for this equation. We define $H(\tau)$ as
\begin{equation}
H(\tau)=\!\int\! f(\mathbf{v})\ln (f(\mathbf{v}))\,d\mathbf{v},  \label{b30}
\end{equation}
and use Eqs.~\eqref{eq1} and (\ref{b20}) to express $dH/d\tau$ as
\begin{equation}
\frac{dH}{d\tau}=-\frac{1}{4}\!\int\! \Psi [f]T(\mathbf{v}_{A},\mathbf{v}%
_{B}; \mathbf{v}_{C},\mathbf{v})d\mathbf{v}_{A}\,d\mathbf{v}_{B}\,d\mathbf{v}%
_{C}\,d \mathbf{v},  \label{b40}
\end{equation}
where
\begin{equation}
\Psi [f]=\big[ f(\mathbf{v}_{A})f(\mathbf{v}_{B})-f(\mathbf{v})f(\mathbf{v}%
_{C})\big] \big[ \ln f(\mathbf{v}_{A})f(\mathbf{v}_{B})-\ln f(\mathbf{v}) f(%
\mathbf{v}_{C})\big] .  \label{b45}
\end{equation}
The function $\Psi [f]$ can be shown to be always nonnegative. We argue that
$(x-y)(\ln x-\ln y)$ is nonnegative for all positive $x$ and $y$; $\ln x$ is
an increasing function and thus $x-y$ and $\ln x-\ln y$ always have the same
sign. Their product is always either positive or zero and zero occurs for $%
x=y$. $T(\mathbf{v}_{A},\mathbf{v}_{B}; \mathbf{v}_{C},\mathbf{v})$ is
intrinsically positive. Therefore the integrand is positive and $dH/d\tau $
is negative.

Following the usual arguments of the H-theorem, the decrease of $H$ stops
only when
\begin{equation}
\ln f(\mathbf{v}_{A})+\ln f(\mathbf{v}_{B})=\ln f(\mathbf{v})+\ln f(\mathbf{v%
}_{C})  \label{b50}
\end{equation}
is satisfied, which implies that $\ln f(\mathbf{v})$ is a collision
invariant. If we choose $T(\mathbf{v}_{A},\mathbf{v}_{B};\mathbf{v}_{C},%
\mathbf{v})$ such that the total momentum and energy is conserved in each
collision, then $\ln f(\mathbf{v})$ must be expressible as a linear
combination of these collision invariants as
\begin{equation}
\ln f(\mathbf{v})=\frac{m}{2\Theta }(\mathbf{v}-\mathbf{v}_{0})^{2}+%
\mbox{constant},  \label{b60}
\end{equation}
where $\Theta $ is the temperature in energy units ($k_{B}=1$) and $m$ is
the mass of a molecule. Here $\mathbf{v}_{0}$ is the velocity of the center
of mass of the system. Hence we have shown that the system approaches the
Maxwell-Boltzmann distribution.

\subsection{Structure of $T(\mathbf{v}_{A},\mathbf{v}_{B};\mathbf{v}_{C}
\mathbf{v})$, and connection with the Boltzmann equation}

We define new variables
\begin{subequations}
\begin{align}
\vv_{T} &=(\vv_{A}+\vv_{B})/2,\quad \uv=\vv_{A}-\vv_{B},\quad u=|
\uv| \\
\vv_{T}^{\prime}& =(\vv+\vv_{C})/2,\quad \uv^{\prime}=
\vv_{C}-\vv,\quad u^{\prime}=| \uv^{\prime}|,
\label{c9}
\end{align}
\end{subequations}
where $\mathbf{v}_{T}$ and $\mathbf{v}_{T}^{\prime }$ are the center of mass
velocities before and after the collision. The Jacobian of the
transformation is unity and integrations can be written in terms of the new
variables. Momentum conservation is imposed on $T(\mathbf{v}_{A},\mathbf{v}%
_{B};\mathbf{v}_{C},\mathbf{v})$ as
\begin{equation}
T(\mathbf{v}_{A},\mathbf{v}_{B};\mathbf{v}_{C},\mathbf{v})=\delta ^{3}(%
\mathbf{v}_{T}-\mathbf{v}_{T}^{\prime })G(\mathbf{u},\mathbf{u}^{\prime }).
\label{c20}
\end{equation}
The integral in Eq.~(\ref{a130}) is then written as
\begin{subequations}
\begin{align}
I &= \!\int\!
f(\vv_{A})f(\vv_{B})T(\vv_{A},\vv_{B};
\vv_{C},\vv)\,d\vv_{A}\,d\vv_{B}\,d\vv_{C}
\label{c30} \\
&= \!\int\! f(\vv+\frac{\uv^{\prime} + \uv}{2})f(\vv-
\frac{\uv-\uv^{\prime}}{2})G(\uv,\uv^{\prime})\,d\uv\,d\uv^{\prime}.
\end{align}
\end{subequations}
The conditions in Eqs.~(\ref{eq1}) and (\ref{a40}) become for $G(\mathbf{u},%
\mathbf{u}^{\prime })$:
\begin{align}
\int \!G(\mathbf{u},\mathbf{u}^{\prime })d\mathbf{u}& =\!\int \!G(\mathbf{u},%
\mathbf{u}^{\prime })d\mathbf{u}^{\prime }=1,  \label{c40} \\
G(\mathbf{u},\mathbf{u}^{\prime })& =G(\mathbf{u}^{\prime },\mathbf{u}).
\label{c50}
\end{align}

Energy conservation requires that $u = u^{\prime}$. If we define unit
vectors $\hat{\mathbf{u}}=\mathbf{u}/u$ and $\hat{\mathbf{n}} =\mathbf{u}%
^{\prime}/u$ and the angle $\theta$ between them as ($\cos \theta =\hat{%
\mathbf{u}} \cdot \hat{\mathbf{n}})$, we can write $G(\mathbf{u},\mathbf{u}%
^{\prime})$ as
\begin{equation}
G(\mathbf{u},\mathbf{u}^{\prime})=\frac{\delta (u^{\prime}-u)}{u^{2}}
g(\theta,u).  \label{c60}
\end{equation}
The condition in Eq.~(\ref{c40}) becomes for $g(\theta,u)$:
\begin{equation}
\int\! g(\theta,u)d\hat{\mathbf{n}}=1,  \label{c70}
\end{equation}
where $g(\theta,u)d\hat{\mathbf{n}}$ is the probability of scattering into
the solid angle $d\hat{\mathbf{n}}$ in the center of mass frame. Then the
integral $I$ in Eq.~(\ref{c30}) becomes
\begin{equation}
I=\!\int\!f(\mathbf{v}+\frac{\mathbf{u}}{2}+u\hat{\mathbf{n}}/2)f(\mathbf{v}
-\frac{\mathbf{u}}{2}+u\hat{\mathbf{n}}/2)g(\theta,u)\,d\mathbf{u}\,d\hat{%
\mathbf{n}}.  \label{c80}
\end{equation}
If we write the $f(\mathbf{v})$ term in Eq.~(\ref{a130}) as
\begin{equation}
f(\mathbf{v})=\!\int\! f(\mathbf{v})f(\mathbf{v} -\mathbf{u})g(\theta,u)\,d%
\mathbf{u}\,d\hat{\mathbf{n}},  \label{c90}
\end{equation}
which follows from Eq.~(\ref{c70}) and the normalization of $f(\mathbf{v})$,
we can write Eq.~(\ref{a130}) as
\begin{equation}
\frac{\partial f(\mathbf{v})}{\partial \tau}=\!\int\! \left[ f(\mathbf{v}%
_{A})f(\mathbf{v}_{B})-f(\mathbf{v})f(\mathbf{v}-\mathbf{u})\right]
g(\theta,u)d \mathbf{u} d\hat{\mathbf{n}},  \label{c100}
\end{equation}
where
\begin{subequations}
\label{eq28}
\begin{align}
\vv_{A} &= \vv+\frac{\uv}{2}+u\nhat/2
\label{c101} \\
\vv_{B} &= \vv-\frac{\uv}{2}+u\nhat/2,
\label{c102}
\end{align}
\end{subequations}
Equation~\eqref{c100} is almost in the form of the Boltzmann equation.

The Boltzmann equation represents a dilute gas for which the collision
probability is proportional to $u\sigma _{T}(u)$, where $\sigma _{T}(u)$ is
the total cross section. We consider a large enough number $R$ such that the
ratio $u\sigma _{T}(u)/R$ for a selected pair is almost always less than
unity. For $\sigma _{T}(u)=\sigma _{0}$ the constant $R/\sigma _{0}$ can be
chosen to be a few (say five) times the rms velocity. Then when a pair is
selected, we take a random number $r$ and allow the collision to occur if $%
r<u\sigma _{T}(u)/R$; we select another pair if $r>u\sigma _{T}(u)/R$.
Although this procedure insures that the collision probability is
proportional to $u\sigma _{T}(u)$, it appears to violate the condition in
Eq.~(\ref{c70}) that all the selected pairs have a collision. To satisfy the
condition in Eq.~(\ref{c70}) we select $g(\theta ,u)$ as
\begin{equation}
g(\theta ,u)=\frac{u\sigma (\theta ,u)}{R}+\Big(1-\frac{u\sigma _{T}(u)}{R}%
\Big) \delta (\hat{\mathbf{u}}-\hat{\mathbf{n}}),  \label{c110}
\end{equation}
where $\sigma (\theta ,u)$ is the differential cross section. The latter is
related to the total cross section $\sigma _{T}(u)$ by
\begin{equation}
\sigma _{T}(u)=\!\int \!\sigma (\theta ,u)\,d\hat{\mathbf{n}}=2\pi \!\int
\!\sigma (\theta ,u)\sin (\theta )d\mathbf{\theta }.  \label{c120}
\end{equation}
The second term in Eq.~(\ref{c110}) transfers the initial velocities to the
final velocities with the probability $1-u\sigma _{T}(u)/R$ and the
collision becomes a null collision. The $\delta (\hat{\mathbf{u}}-\hat{%
\mathbf{n}})$ requires $\hat{\mathbf{u}}=\hat{\mathbf{n}}$ which implies $%
\mathbf{u}^{\prime }=\mathbf{u}$ since $u^{\prime }=u$ from the energy
conservation and $\mathbf{u}^{\prime }=u^{\prime }\hat{\mathbf{n}}$, $%
\mathbf{u=}u\hat{\mathbf{u}}$. This means $\mathbf{v}_{A}-\mathbf{v}_{B}=%
\mathbf{v}_{C}-\mathbf{v}.$ We also have $\mathbf{v}_{A}+\mathbf{v}_{B}=%
\mathbf{v}_{C}+\mathbf{v}$ from center of mass velocity conservation. These
two equations yield $\mathbf{v}_{C}=\mathbf{v}_{A}$ and $\mathbf{v}=\mathbf{v%
}_{B}$ and velocities have not changed. A normal collision occurs with the
probability $u\sigma _{T}(u)/R$. It is easy to verify that $g(\theta ,u)$
given in Eq.~(\ref{c110}) satisfies the condition in Eq.~(\ref{c70}).

If we substitute $g(\theta ,u)$ in Eq.~(\ref{c110}) into Eq.~(\ref{c100}),
we obtain
\begin{equation}
\frac{\partial f(\mathbf{v})}{\partial (\tau /R)}=\!\int \![f(\mathbf{v}%
_{A})f(\mathbf{v}_{B})-f(\mathbf{v})f(\mathbf{v}-\mathbf{u})]u\sigma (\theta
,u)\,d\mathbf{u}\,d\hat{\mathbf{n}},  \label{c130}
\end{equation}
where $\mathbf{v}_{A}$ and $\mathbf{v}_{B}$ were given in Eq.~(\ref{eq28}).
Equation~\eqref{c130} is essentially the Boltzmann equation with the
difference that $f(\mathbf{v})$ is the probability density in velocity space
whereas the Boltzmann equation is written in terms of the probability
density in both physical and velocity space. If the volume of the cell
containing the molecules is $V$, then we can write Eq.~\eqref{c130} for $F(%
\mathbf{v})=(N/V)f(\mathbf{v})$ as
\begin{equation}
\frac{\partial F(\mathbf{v})}{\partial t}=\!\int \!\big[ F(\mathbf{v}_{A})F(%
\mathbf{v}_{B})-F(\mathbf{v})F(\mathbf{v}-\mathbf{u})\big] u\sigma (\theta
,u)\,d\mathbf{u}\,d\hat{\mathbf{n}},  \label{eq32}
\end{equation}
where $t=\tau V/RN=2nV/RN^{2}$ is interpreted as the physical time. Equation~%
\eqref{eq32} is the Boltzmann equation for a homogeneous gas.

\section{Discussion}

Let us summarize the direct simulation Monte Carlo algorithm for solving the
Boltzmann equation. We choose a sufficiently large $R$ such that only a
negligible fraction of the selected pairs (say less than one in a thousand)
violate the condition $u\sigma _{T}(u)/R\leq 1$. Then we select pairs
randomly and let them collide with probability $u\sigma _{T}(u)/R$. The
latter is achieved by generating a random number $r$ and letting the
collision occur if $r\leq u\sigma _{T}(u)/R$. If a pair collides, then in
the center of mass system the collision occurs within the solid angle $d\hat{%
\mathbf{n}}$ with probability $P(\theta )d\hat{\mathbf{n}}=[\sigma (\theta
,u)/\sigma _{T}(u)]d\hat{\mathbf{n}}$. Suppose that we put the $z$-axis
along $\mathbf{u}$ and we need to determine $\hat{\mathbf{n}}=\mathbf{u}%
^{\prime }/u$, which is determined by the angles $\theta $ and $\phi $. To
determine $\theta $ we need to generate a random value of $\theta $ by
converting the random numbers produced by a uniform probability distribution
to random numbers in the interval $(0,\pi )$ according to the probability
distribution $P(\theta )$. The $\phi $ angles in the interval $(0,2\pi )$
are equally likely. In this way we determine the final velocities of the
particles as $\mathbf{u}^{\prime }$ and $-\mathbf{u}^{\prime }$ in the
center of mass frame. By adding the center of mass velocity we find the
final velocities in the lab frame. After storing the final velocities of the
particles, we choose another pair and repeat the same process. The physical
time is $t=2nV/N^{2}R$, where $n$ is the number pairs chosen to make
attempts for a collision. If the number of collisions in a given time is
required, we can count the successful attempts for a collision. In Ref.~%
\onlinecite{Bird94} this algorithm for keeping track of the time is called
the ``no time counter method.''

The original method of Bird\cite{Bird94} to keep track of the time was the
time counter method. Consider a narrow interval of $u\sigma _{T}(u)$ values.
For a large $n$ there will be $\Delta n$ pairs with $u\sigma _{T}(u)$ values
in this interval. Of these, only $(u\sigma _{T}(u)/R)\Delta n$ of them will
make collisions corresponding to a time interval $(2V/N^{2}R)\Delta n$. Thus
the elapsed time per successful attempt is
\begin{equation}
\Delta t=\frac{(2V/N^{2}R)\Delta n}{(u\sigma _{T}(u)/R)\Delta n}=\frac{2V}{%
N^{2}u\sigma _{T}(u)}.  \label{tc}
\end{equation}
In the time counter method we let every pair collide, increase time by $%
\Delta t$ ($t\rightarrow t+\Delta t$) after each collision, and keep
selecting pairs and colliding them until we reach the desired time. Every
collision will cause a different time increment depending on the value of $%
u\sigma _{T}(u)$. One disadvantage of this method is that if a collision
with a low $u\sigma _{T}(u)$ occurs, the time increment will be large. Such
collisions can occur with pairs having almost equal velocities. The time
counter method was declared ``obsolete' in Ref.~\onlinecite{Bird94}. But it
is useful to be aware of the method since it is widely used in the past and
it might come across in some papers.

If the purpose of the simulation is to demonstrate that the velocity
distribution approaches the Maxwell-Boltzmann distribution, we could let all
the selected pairs make a collision and the velocity distribution will
converge to the Maxwell-Boltzmann distribution. This simplification
corresponds to $u\sigma_{T}(u)/R=1$ or $\sigma_{T}(u)=R/u$, where the total
cross section is inversely proportional to the relative velocity. Also, if
it is desired to not discuss cross sections and the time tracking method, it
is convenient to assume isotropic scattering in the center of mass frame.
Then $\mathbf{u}^{\prime}$ can be calculated by taking a random unit vector $%
\hat{\mathbf{n}}$ and multiplying it by $u$. These two simplifications make
the programming easier and an undergraduate student with some programming
background can write a program demonstrating the Maxwell-Boltzmann
distribution.

Direct simulation methods are also applicable to radiative processes and
chemical reactions and the present formalism generalizes to all these cases
in a more or less straightforward fashion for homogeneous gases. Such
generalizations can be a useful teaching tool and a fertile field for
student projects.


\begin{thebibliography}{99}
\bibitem{Bird70}  G. A. Bird, ``Direct simulation and the Boltzmann
equation,'' Phys. Fluids \textbf{13} (11), 2676--2681 (1970).

\bibitem{Garcia}  F.J. Alexander and A.L. Garcia, ``The direct simulation
Monte Carlo method,'' Computers in Physics, 11 (6), 588-593 (1997)

\bibitem{Bird94}  G. A. Bird, \textit{Molecular Gas Dynamics and the Direct
Simulation of Gas Flows} (Clarendon Press, Oxford, 1994).

\bibitem{Carlo}  C. Cercignani, \textit{Rarified Gas Dynamics: From Basic
Concepts to Actual Calculations} (Cambridge University Press, 2000)

\bibitem{Shen}  C. Shen, \textit{Rarified Gas Dynamics: Fundamentals,
Simulations and Micro Flows} (Springer-Verlag, Berlin Heidelberg, 2005)

\bibitem{Nanbu80}  K. Nanbu, ``Direct simulation scheme derived from the
Boltzmann equation. I. Monocomponent gases,'' J. Phys. Soc. Japan \textbf{49}
(5), 2042-2049 (1980).

\bibitem{Babovsky1}  H. Babovsky, ``A convergence proof for Nanbu's
Boltzmann simulation scheme,'' Eur. J. Mech. B/Fluids \textbf{8} (1), 41-55
(1989).

\bibitem{Babovsky2}  H. Babovsky and R. Illner, ``A convergence proof for
Nanbu's simulation method for the full Boltzmann equation,'' SIAM J. Numer.
Anal. \textbf{26} (1), 45-65 (1989).

\bibitem{Wagner92}  W. Wagner, ``A convergence proof for Bird's direct
simulation Monte Carlo method for the Boltzmann equation,'' J. Stat. Phys.
\textbf{66} (3/4), 1011-1044 (1992).

\bibitem{Novak70}  J. Novak and A. B. Bortz, ``The evolution of
two-dimensional Maxwell-Boltzmann distribution,'' Am. J. Phys. \textbf{38}
(12), 1402-1406 (1970).

\bibitem{Eger82}  M. Eger and M. Kress, ``Simulation of Boltzmann processes:
An energy space model,'' Am. J. Phys. \textbf{50} (2), 120-124 (1982).

\bibitem{Bonomo84}  R. P. Bonomo and F. Riggi, ``The evolution of the speed
distribution for a two dimensional ideal gas: A computer simulation,'' Am.
J. Phys. \textbf{52} (1), 54-55 (1984).

\bibitem{Sauer81}  G. Sauer, ``Teaching classical statistical mechanics: A
simulation approach,'' Am. J. Phys. \textbf{49} (1), 13-19 (1981).

\bibitem{Berger88}  J. Berger, ``Kinetic illustrations for thermalization,''
Am. J. Phys. \textbf{56} (10), 923-928 (1988).

\bibitem{Huang}  K. Huang, \textit{Statistical Mechanics} (John Wiley \&
Sons, 1987), 2nd ed., Chap. 3.
\end{thebibliography}
\end{document}